\documentclass[conference]{IEEEtran}
\IEEEoverridecommandlockouts

\usepackage{amsmath,amssymb,amsfonts,amsthm}
\usepackage{mathtools}
\usepackage{algorithmic}
\usepackage{graphicx}
\usepackage{textcomp}
\usepackage{xcolor}
\def\BibTeX{{\rm B\kern-.05em{\sc i\kern-.025em b}\kern-.08em
    T\kern-.1667em\lower.7ex\hbox{E}\kern-.125emX}}

\usepackage[english]{babel}
\usepackage[T1]{fontenc}
\usepackage[cmintegrals]{newtxmath}
\usepackage{bm}
\usepackage[caption=false,font=footnotesize,farskip=0pt]{subfig}
\usepackage{multirow}
\usepackage[ruled,vlined,linesnumbered]{algorithm2e}
\usepackage{enumitem}
\usepackage{soul}
\usepackage[normalem]{ulem}
\usepackage{booktabs}
\usepackage{pgfplots}
\usepackage{balance}

\usepackage[backend=biber, style=ieee, maxnames=2, minnames=1, doi=false, url=false]{biblatex}
\usepackage{csquotes}
\usepackage[acronym]{glossaries}
\newacronym{3gpp}{3GPP}{3rd Generation Partnership Project}
\newacronym{5g}{5G}{the 5th generation of mobile networks}
\newacronym{6g}{6G}{sixth generation of mobile networks}
\newacronym{awgn}{AWGN}{additive white Gaussian noise}
\newacronym{b5g}{B5G}{5G and Beyond}
\newacronym{ber}{BER}{bit error rate}
\newacronym{bler}{BLER}{block error rate}
\newacronym{bs}{BS}{base station}
\newacronym{cdf}{CDF}{cumulative distribution function}
\newacronym{cml}{CML}{commercial microwave link}
\newacronym{crb}{CRB}{Cramér-Rao bound}
\newacronym{csi}{CSI}{channel state information}
\newacronym{dl}{DL}{downlink}
\newacronym{dmrab}{DMRAB}{disjoint matching and resource allocation benchmark}
\newacronym{embb}{eMBB}{Enhanced Mobile Broadband}
\newacronym{gap}{GAP}{Generalized Assignment Problem}
\newacronym{geo}{GEO}{Geosynchronous Earth Orbit}
\newacronym{gsl}{GSL}{Ground-to-Satellite Link}
\newacronym{iot}{IoT}{Internet of Things}
\newacronym{isac}{ISAC}{integrated sensing and communication}
\newacronym{isl}{ISL}{Inter-Satellite Link}
\newacronym{jmra}{JMRA}{joint matching and resource allocation}
\newacronym{kpi}{KPI}{key performance indicator}
\newacronym{leo}{LEO}{Low Earth Orbit}
\newacronym{mac}{MAC}{medium access control}
\newacronym{mcs}{MCS}{modulation and coding scheme}
\newacronym{mgap}{MGAP}{Multi-Level Generalized Assignment Problem}
\newacronym{milp}{MILP}{mixed-integer linear problem}
\newacronym{mimo}{MIMO}{multiple-input multiple-output}
\newacronym{ml}{ML}{machine learning}
\newacronym{mle}{MLE}{maximum likelihood estimator}
\newacronym{mr}{MR}{maximum ratio}
\newacronym{mse}{MSE}{mean-squared error}
\newacronym{nmse}{NMSE}{normalized mean-squared error}
\newacronym{noma}{NOMA}{non-orthogonal multiple access}
\newacronym{nr}{NR}{New Radio}
\newacronym{ntn}{NTN}{Non-Terrestrial Network}
\newacronym{ofdma}{OFDMA}{Orthogonal Frequency-Division Multiple Access}
\newacronym{pdf}{PDF}{probability density function}
\newacronym{pdv}{PDV}{packet delay variation}
\newacronym{pmf}{PMF}{probability mass function}
\newacronym{ppp}{PPP}{Poisson point process}
\newacronym{qos}{QoS}{quality of service}
\newacronym{ra}{RA}{resource allocation}
\newacronym{ran}{RAN}{radio access network}
\newacronym{rcs}{RCS}{radar cross-section}
\newacronym{rssi}{RSSI}{received signal strength indicator}
\newacronym{rtt}{RTT}{round-trip time}
\newacronym{rv}{RV}{random variable}
\newacronym{snr}{SNR}{signal-to-noise ratio}
\newacronym{sinr}{SINR}{signal-to-interference-plus-noise ratio}
\newacronym{ss}{SS}{synchronization signal}
\newacronym{tdd}{TDD}{time-division duplexing}
\newacronym{twi}{TWI}{temporal window of integration}
\newacronym{uav}{UAV}{unmanned aerial vehicle}
\newacronym[longplural=users equipment]{ue}{UE}{user equipment}
\newacronym{ul}{UL}{uplink}
\newacronym{ula}{ULA}{uniform linear array}

\newtheorem{definition}{Definition}
\newtheorem{assumption}{Assumption}

\usepackage{stfloats}
\fnbelowfloat

\definecolor{amaranth}{rgb}{0.9, 0.17, 0.31}

\usepgfplotslibrary{fillbetween}

\pgfplotsset{
    compat=1.18,
    tick label style={font=\scriptsize},
    label style={font=\scriptsize},
    legend style={font=\scriptsize,draw=none,row sep=-2pt,inner sep=0,outer sep=0,fill=none},
    tick style={color=black},
    major tick length=3pt,
    minor tick length=1.5pt,
    label shift=-4pt
}

\definecolor{xgfs_normal6_blue}{RGB}{64, 83, 211}
\definecolor{xgfs_normal6_yellow}{RGB}{221, 179, 16}
\definecolor{xgfs_normal6_red}{RGB}{181, 29, 20}
\definecolor{xgfs_normal6_lightblue}{RGB}{0, 190, 255}
\definecolor{xgfs_normal6_pink}{RGB}{251, 73, 176}
\definecolor{xgfs_normal6_green}{RGB}{0, 178, 93}
\definecolor{xgfs_normal6_gray}{RGB}{202, 202, 202}

\begin{document}

\title{Medium Access for Multi-Cell ISAC Through Scheduling of Radar and Communication Tasks
\thanks{The work by J. H. Inacio de Souza and P. Popovski was supported by the Villum Investigator Grant “WATER” from the Velux Foundation, Denmark. The work by F. Saggese was supported by the Horizon Europe MSCA Postdoctoral Fellowships with Grant~101204088. The work by K. Chen-Hu was supported by the Ramón y Cajal Programme under Grant RYC2024-048736-I funded by MICIU/AEI/10.13039/501100011033 and FSE+.}}

\author{\IEEEauthorblockN{João Henrique Inacio de Souza\IEEEauthorrefmark{1}, Fabio Saggese\IEEEauthorrefmark{2}, Kun Chen-Hu\IEEEauthorrefmark{3}, Petar Popovski\IEEEauthorrefmark{1}}
\IEEEauthorblockA{\IEEEauthorrefmark{1}\textit{Department of Electronic Systems}, \textit{Aalborg University}, Denmark. E-mail: \{
\pdfstartlink
    attr{/Border [0 0 0]}
    user{/Subtype /Link /A << /S /URI /URI (mailto:jhids@es.aau.dk) >>}%
    jhids%
\pdfendlink
,%
\pdfstartlink
    attr{/Border [0 0 0]}
    user{/Subtype /Link /A << /S /URI /URI (mailto:petarp@es.aau.dk) >>}%
    petarp%
\pdfendlink
\}@es.aau.dk}
\IEEEauthorblockA{\IEEEauthorrefmark{2}\textit{Department of Information Engineering}, \textit{University of Pisa}, Italy. E-mail: fabio.saggese@ing.unipi.it}
\IEEEauthorblockA{\IEEEauthorrefmark{3}\textit{Department of Signal Theory and Communications}, \textit{Universidad de Alcal\'a}, Spain. E-mail: kun.chen@uah.es}
}

\maketitle

\begin{abstract}
This paper focuses on communication, radar search, and tracking task scheduling in multi-cell integrated sensing and communication~(ISAC) networks under quality-of-service constraints. We propose a medium access control framework that multiplexes these tasks while optimizing radar scan patterns through an interference-aware scheduling algorithm. Specifically, the proposed framework employs time-domain task scheduling and beam selection, formulated as an assignment problem, to mitigate inter-task and inter-cell interference, respectively. Simulations show that our solution guarantees target communication throughput, sensing target detection probability, and sensing signal-to-interference-plus-noise ratio with improved resource efficiency over baseline schemes, highlighting the benefits of coordinated scheduling in multi-cell ISAC.
\end{abstract}

\begin{IEEEkeywords}
    Integrated sensing and communication, space-division multiple access, resource management, interference.
\end{IEEEkeywords}

%
%
\section{Introduction}
\label{sec:into}

The vision for the \gls{6g} centers on \emph{connected intelligence}, by, among other solutions, seamlessly blending communication and localization~\cite{Liu2022}. This redefines the traditional role of a \gls{bs}, as it now serves both to support digital links as well as interface the physical world through radar probes. By unifying these functions into one transmitter–receiver platform, \gls{isac} eliminates redundant infrastructure and can achieve tighter performance synergies than independent radar and communication systems~\cite{Meng2024cooperativeISAC}.

Most \gls{isac} research to date has focused on waveform design to guarantee a certain \gls{qos} in terms of communication throughput and sensing metrics, such as range resolution, Doppler tolerance, and angular discrimination~\cite{Wei2023isacsignals}. Efforts at higher layers have looked at joint resource allocation strategies---time‑frequency partitioning, multiple access, and power control---that balance the conflicting demands of radar and data traffic in single-cell environments, \emph{e.g.},~\cite{Dong2023unifiedra, Cao2024isacra}. A more recent strand of work has introduced the notion of a cooperative \gls{isac} network, in which multiple sensing‑capable \glspl{bs} coordinate their transmissions and jointly process echoes~\cite{Meng2024cooperativeISAC}. Such an approach can significantly improve detection coverage, suppress mutual interference, and refine localization accuracy by fusing observations from diverse viewpoints~\cite{Buzzi2024cellfreeISAC, Li2024multistaticISAC}.

A viable approach enabling cooperative \gls{isac} is exploiting the current multi-cell infrastructure. Localization performance typically benefits from wide bandwidths, which, under conventional cellular planning, implies full frequency reuse across neighboring cells. Unfortunately, this also amplifies radar‑to‑radar interference, which in turn may eventually degrade radar performance instead of enhancing it~\cite{Munari2018}. What is thus needed is a space‑time spectrum sharing strategy that orchestrates when and where each \gls{bs} performs sensing and communication, so as to localize radar targets without sacrificing \glspl{ue} \gls{qos}.
On this line, recent works have focused on beamforming design for \gls{mimo} \glspl{bs}. The authors of~\cite{Li2024multistaticISAC} and~\cite{Wang2024interference-multicell} proposed beamforming optimization to maximize sensing coverage in a multi-\gls{bs}, single-cell environment, and to mitigate inter-cell interference in multi-cell \gls{isac} systems, respectively. \cite{Babu2024multicell} focuses on a multi-\gls{bs} beamforming design able to exploit shared \gls{csi} and data among \glspl{bs} to enable bistatic sensing between cells.
\cite{Ma2025multicell}~extends the beamforming design to multi-array \glspl{bs} and transmit/receive mode selection, maximizing the sensing \gls{sinr} subject to communication \gls{sinr} constraints.

\begin{figure}[bt]
    \centering
    \includegraphics[width=\columnwidth]{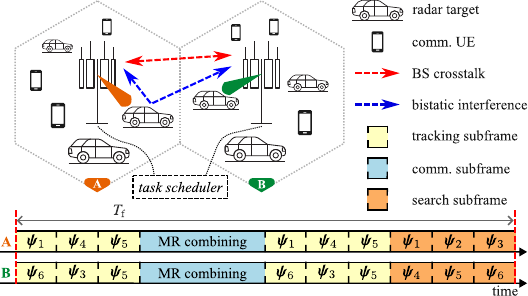}
    \caption{Multi-cell sensing and communication network and proposed frame design for task coexistence. During the tracking and search subframes, \glspl{bs} transmit radar pulses with the beamforming vectors $\bm{\psi}_j\in\Psi_i$. During the communication subframe, \glspl{bs} employ \gls{mr} combining to receive \glspl{ue}' data.}
    \label{fig:scenario}
\end{figure}

From the \gls{mac} layer perspective, scheduling for multi-cell \gls{isac} remains underexplored. Moreover, sensing itself entails heterogeneous tasks---\emph{search}, which scans wide areas for new objects, and \emph{tracking}, which allocates resources to monitor known targets with high precision~\cite{Aubry2024}. A practical framework must thus allocate resources not only across cells but also between these sensing modes, while safeguarding minimum data rate guarantees. This challenge reduces to designing a multi-cell \gls{mac} framework that dynamically orchestrates shared space–time resources for communication and radar tasks. In this work, we propose a \gls{mac} layer framework where communication and sensing are time multiplexed, enabling a tractable optimization of search, tracking, and data transmission tasks. Specifically, the proposed framework employs time-domain task scheduling and beam selection---formulated as an assignment problem---to mitigate inter-task and inter-cell interference, respectively. Then, communication and radar tasks are scheduled to satisfy communication throughput, sensing target detection, and sensing \gls{sinr}. Our approach jointly coordinates cells to achieve improved sensing–communication coexistence with efficient resource utilization. Simulations prove that the proposed framework satisfies radar requirements with reliability above 0.999, outperforming the baselines.

%
%
\section{System Model}
\label{sec:system-model}

We analyze a multi-cell scenario comprised of a cluster of $N_\text{c}$ cells with radius $R$, each having a \gls{bs} at its center, as shown in Fig.~\ref{fig:scenario}. The overall bandwidth available for the cluster is $W$, while the carrier wavelength is denoted by $\lambda$. The \glspl{bs} are equipped with \glspl{ula} of $N_\text{a}$ half-wavelength spaced antennas lying in the $xy$-plane (parallel to the ground). In each cell, the \gls{bs} aims to perform communication towards \glspl{ue} while performing radar operations, which comprises \emph{search} and \emph{tracking} tasks, further described in Sec.~\ref{sec:radar-task}. Finally, communication and radar tasks are assumed to be multiplexed in the time domain, as detailed in Sec.~\ref{sec:rad-comm}, to neglect the inter-task interference generated and study the fundamental trade-offs between communication, search, and tracking.

\paragraph{Communication model}
We consider the \gls{ul} direction of a multi-cell massive \gls{mimo} system employing frequency reuse factor 1, \emph{i.e.}, all \glspl{ue} share the same frequency resources~\cite{Bjornson2017}. \gls{bs}~$i$ has a set $\mathcal{L}_i$ of \glspl{ue} to serve. At the start of the communication operations, the \gls{csi} is acquired through standard pilot-based methods. To present a simple communication metric for the trade-off analysis, we assume uncorrelated Rayleigh fading, no occurrence of pilot contamination, and adoption of \gls{mr} combining. Accordingly, the effective \gls{ul} \gls{sinr} of user $l\in\mathcal{L}_i$ in cell $i$ results in~\cite[Theo. 4.4]{Bjornson2017}
\begin{equation} \label{eq:comm_sinr}
    \gamma_{il} = \frac{N_\text{a} p_\text{c} \beta_{il}^i}{N_0W + \sum_{j=1}^{N_\text{c}} \sum_{k\in\mathcal{L}_j} p_\text{c} \beta_{jk}^i - p_\text{c} \beta_{il}^i},
\end{equation}
where $p_\text{c}$ is the transmit power, $\beta_{jk}^i$ is the large-scale fading coefficient of the $k$-th \gls{ue} of cell $j$ to \gls{bs} $i$, and $N_0$ is the noise power spectral density.

\paragraph{Radar model}
All cells in the system employ the full available system bandwidth $W$ for radar operations, generating inter-cell radar interference~\cite{Liu2022}. Let us evaluate the average \gls{sinr} of the echo reception considering a single radar pulse transmission with power $p_\text{r}$ and a generic target~$k$, which position is described by the range $\rho_{ik}$ and azimuth $\theta_{ik}\in[-\pi,\pi)$, both defined w.r.t. the position of \gls{bs} $i$. To characterize the target reflectivity, we consider a Swerling~1 model with $\sigma_{ij}$ being the target \gls{rcs} when \gls{bs}~$j$ transmits and the \gls{bs}~$i$ receives. For $i = j$, the $\sigma_{ij}$ is the monostatic \gls{rcs}, while it is bistatic for $i\neq j$.
Moreover, the following assumption holds.

\begin{assumption}\label{assumption:gaussian-sources}
    The impinging signals in the radar receiver are modeled as independent, memoryless Gaussian sources.
\end{assumption}

The \glspl{bs} use beamforming at their \glspl{ula} for pulse transmission and echo reception. In this regard, $\bm{\phi}_i\in\mathbb{C}^{N_\text{a}}$ is the beamforming vector of \gls{bs}~$i$. We further denote by $\Psi_i=\{\bm{\psi}_j\}_{j=1}^{N_\text{l}}$, with $\bm{\psi}_j \in \mathbb{C}^{N_\text{a}}$ and $N_\text{l}=|\Psi_i|$, the \emph{beamforming codebook}, \emph{i.e.}, the set of beamforming vectors available to \gls{bs}~$i$, designed to provide the minimum number of configurations able to probe the whole coverage area with sufficient power, see, \emph{e.g.},~\cite{Babu2024multicell}. Assuming far-field propagation for all the points in space relevant to the analysis, we denote as $G(\theta; \bm{\phi}_i)$ the array power pattern for a signal traveling to or from azimuth angle $\theta$ w.r.t. \gls{bs}~$i$ loading beamforming vector $\bm{\phi}_i\in\Psi_i$.

From the \emph{radar range equation}, the average received power by \gls{bs}~$i$ resulted from the scattering from target~$k$ while employing beamformer $\bm{\phi}_i$ is~\cite{Richards2005}:
\begin{equation}\label{eq:monostatic-power}
    M_{ik}(\bm{\phi}_i)=\frac{p_\text{r}G(\theta_{ik}; \bm{\phi}_i)^2\lambda^2\sigma_{ik}}{(4\pi)^3\rho^4_{ik}},
\end{equation}
referred to as the received \emph{monostatic power}.

The interference provoked by \gls{bs}~$j$ on \gls{bs}~$i\neq j$ can be described by two components. The first one arises when the transmit and receive array power gain of the two \glspl{bs} are simultaneously nonzero, \emph{i.e.}, $G(\theta_{ik}; \bm{\phi}_i)\neq 0$ and $G(\theta_{jk}; \bm{\phi}_j)\neq 0$ for some $k$. This component is referred to as the received \emph{bistatic interference power}, defined as~\cite{Richards2005}:
\begin{equation}\label{eq:bistatic-power}
    B_{ijk}(\bm{\phi}_i,\bm{\phi}_j) = \frac{p_\text{r}G(\theta_{jk}; \bm{\phi}_j) G(\theta_{ik}; \bm{\phi}_i) \lambda^2\sigma_{ij}}{(4\pi)^3\rho^2_{jk}\rho^2_{ik}}.
\end{equation}
The second component arises when the transmit and receive array power gain of the two \glspl{bs} are simultaneously nonzero for the directions of each other, \emph{i.e.}, $G(\vartheta_{ij}; \bm{\phi}_i)\neq0$ and $G(\vartheta_{ji}; \bm{\phi}_j)\neq0$, where $\vartheta_{ij}\in[-\pi,\pi)$ is the azimuth of \gls{bs}~$j$ w.r.t. the position of \gls{bs}~$i$, and vice versa. We name it the received \emph{\gls{bs} crosstalk power}, resulting in:
\begin{equation}\label{eq:bs-crosstalk-power}
    C_{ij}(\bm{\phi}_i,\bm{\phi}_j)=\frac{p_\text{r}G(\vartheta_{ij}; \bm{\phi}_i)G(\vartheta_{ji}; \bm{\phi}_j)\lambda^2}{(4\pi)^2\varrho^2_{ij}},
\end{equation}
where $\varrho_{ij}$ is the distance between \gls{bs} $j$ and \gls{bs} $i$.

We define the set $\mathcal{K}$ containing the indices of the targets in the area covered by the cluster. Additionally, let $\bm{\phi}=[\bm{\phi}_1^\transp~\cdots~\bm{\phi}_{N_\text{c}}^\transp]^\transp$ be the vector collecting the beamformers set by all \glspl{bs}. Thus, using eqs.~\eqref{eq:monostatic-power},~\eqref{eq:bistatic-power}, and~\eqref{eq:bs-crosstalk-power}, we obtain the received \gls{sinr} of target $k$'s echo as a function of $\bm{\phi}$ as in eq.~\eqref{eq:sinr} at the top of the page, where $N_0W$ is the noise power.\footnote{The effect of clutter and reflectivity of \glspl{ue} is neglected for simplicity of analysis and will be included in future works.} The received communication \gls{sinr} in~\eqref{eq:comm_sinr} and radar \gls{sinr} in~\eqref{eq:sinr} are used to evaluate the communication and radar performances. 

\begin{figure*}
    \begin{equation}
        \label{eq:sinr}
        \Gamma_{ik}(\bm{\phi})= M_{ik}(\bm{\phi}_i) \Bigg(N_0W+\sum_{\substack{k'\in\mathcal{K} \\ k'\neq k}}M_{ik'}(\bm{\phi}_i) + \sum_{\substack{j=1 \\ j\neq i}}^{N_\text{c}}\sum_{k\in\mathcal{K}}B_{ijk}(\bm{\phi}_i,\bm{\phi}_j)+\sum_{\substack{j=1 \\j\neq i}}^{N_\text{c}}C_{ij}(\bm{\phi}_i, \bm{\phi}_j)\Bigg)^{-1},
    \end{equation}  
    \rule{\textwidth}{0.05em}
\end{figure*}

%
%
\section{Radar and Communication Coexistence}\label{sec:rad-comm}

Here, we describe the communication and radar tasks performed by the network. To perform a space-time resource sharing strategy, the \glspl{bs} operations are synchronized following a shared frame-based structure\footnote{The control signaling burden between the \glspl{bs} for a precise synchronization is left for future studies.} of duration $T_\text{f}$, shown at the bottom of Fig.~\ref{fig:scenario}. In each frame, each task occupies a specific subframe, namely $T_\text{c}$ for communication, $T_\text{s}$ for search, and $T_\text{t}$ for tracking; this results in a total duration of $T_\text{f} = T_\text{c} + T_\text{s} + T_\text{t}$. Each \gls{bs} can execute only a single task at a time.

\subsection{Communication Task}\label{sec:comm-task}

The communication task comprises the reception of \gls{ul} data from the \glspl{ue} in a cell. Each \gls{bs} executes only one communication task per frame, and throughput is the \gls{qos} metric used to evaluate performance.
According to eq.~\eqref{eq:comm_sinr}, the network throughput
is lower bounded by~\cite{Bjornson2017}
\begin{equation}
    \label{eq:throughput}
    S = \frac{T_\text{c}}{T_\text{f}} W \sum_{i=1}^{N_\text{c}} \sum_{l\in\mathcal{L}_i} \log_2\left(1+\gamma_{il}\right).
\end{equation}
Eq.~\eqref{eq:throughput} shows the direct relation between throughput and duration of the communication subframe.

\subsection{Radar Tasks}\label{sec:radar-task}

Each \gls{bs} performs search and tracking radar tasks. The former focuses on the discovery of new targets, while the latter focuses on tracking parameters of already discovered targets.

\paragraph{Search task}
At this stage, the \gls{bs} has no prior information about the potential targets' positions, so it needs to scan the entire cell area to identify them. Each search task configures a \emph{detection problem}, which is defined as a \emph{hypothesis testing} based on the echoes of $N_\text{p}$ radar pulses transmitted when the \glspl{bs} employ beamforming vectors to probe specific directions. To complete a single probe, the transmission of $N_\text{p}$ pulses takes $T_\text{d}$ seconds, that is, $T_\text{d}$ is the radar dwell time.

The search performance is evaluated through the \emph{probability of target detection}. According to Assumption~\ref{assumption:gaussian-sources} and that $N_\text{p}\Gamma_{ik}(\bm{\phi})>1$, the probability of \gls{bs}~$i$ detecting target~$k$ can be approximated as~\cite[eq. (6.99)]{Richards2005}:
\begin{equation}
    \label{eq:prob-detection}
    P^\text{d}_{ik}(\bm{\phi}) = \left( 1+\frac{1}{N_\text{p} \Gamma_{ik}(\bm{\phi})} \right)^{N_\text{p}-1} \exp \left\{ \frac{-\tau}{1+N_\text{p}\Gamma_{ik}(\bm{\phi})} \right\},
\end{equation}
where $\tau$ is the detection threshold, set according to the \emph{probability of false alarm} given by $P^\text{fa} = 1 - I ( \tau/\sqrt{N_\text{p}}, N_\text{p}-1 )$, where $I(\cdot,\cdot)$ denotes the incomplete Gamma function~\cite[eq. (6.80)]{Richards2005}. Eq.~\eqref{eq:prob-detection} shows that the detection probability for a cell depends on the beamforming vectors loaded by all \glspl{bs}, revealing the need for a joint selection of the beamformers from the respective codebooks to ensure a target performance.

To scan the whole network coverage area, the \glspl{bs} need to execute a total of $N_\text{c}N_\text{l}$ searches. Given that, let $D_\text{s} \in [N_\text{l}, N_\text{c} N_\text{l}]$ denote the number of consecutive \emph{dwell intervals} required to execute the task. Specifically, $D_\text{s} \leq N_\text{c}N_\text{l}$ may be achieved through a space-time sharing strategy, allowing \glspl{bs} to execute searches within the same time-frequency resources. Importantly, depending on its duration $T_\text{s}$, the search subframe may not accommodate $D_\text{s}$ dwell intervals, meaning that a single network scan will span multiple frames. The time performance of the search task is evaluated through the \emph{search rate}, given by
\begin{equation}
    \label{eq:search-rate}
    R_\text{s} = \frac{T_\text{s}}{D_\text{s}T_\text{d}}~\text{[full scans/frame]},
\end{equation}
illustrating the relationship between the search subframe duration, the number of dwell intervals, and the radar dwell time.

\paragraph{Tracking task}
The tracking task is performed towards targets that have already been detected to retrieve their sensing parameters, \emph{e.g.}, precise range, Doppler, or angular direction. Since the search task acquired information on approximate target direction, \emph{i.e.}, the one illuminated by the beamformer $\bm{\phi}_i$ for \gls{bs} $i$, the scan of the tracking task can be limited to a smaller set of range bins and spatial angles. Let us assume each \gls{bs} tracks $N_\text{t}$ targets previously detected in its cell through search. Each tracking task obtains a snapshot of the sensing parameters from a single target. Based on a series of snapshots, each \gls{bs} runs a track algorithm on its $N_\text{t}$ targets, \emph{e.g.}, a Kalman filter for tracking the trajectory of targets over time.

The performance of any estimation procedure strongly depends, among other factors, on the \gls{sinr} of the echoes received from the radar pulses~\cite{Liu2022}. For this reason, we adopt the \gls{sinr} in \eqref{eq:sinr} to evaluate tracking performance without defining a specific method. To ensure that the error covariances of the targets' sensing parameters are bounded in the tracking algorithm, the \glspl{bs} execute tracking tasks for each target at a fixed rate $R_\text{t}$, referred to as the \emph{tracking update rate}. Let $D_\text{t}\in[N_\text{t},N_\text{c}N_\text{t}]$ denote the number of tracking dwell intervals required for \glspl{bs} to probe all targets. Then, the duration of the tracking subframe is calculated as
\begin{equation}
    \label{eq:tracking-subframe}
    T_\text{t} = \lfloor T_\text{f}R_\text{t} \rfloor D_\text{t}T_\text{d},
\end{equation}
where $\lfloor T_\text{f}R_\text{t} \rfloor$ indicates the number of times a \gls{bs} needs to revisit a target within a frame. Note that, as $T_\text{t} \leq T_\text{f}$, the number of targets that can be tracked and the rate at which they can be updated are limited. Eq.~\eqref{eq:tracking-subframe} is adopted to evaluate the time performance of the tracking tasks, and links the tracking subframe to the frame duration, the tracking update rate, the number of tracked targets, and the radar dwell time.

%
%
\section{Task Scheduling Strategy}\label{sec:scheduling}

In this section, we introduce a task scheduling strategy with two objectives: 1) find \glspl{bs}' scan patterns, \emph{i.e.}, the selection of beamforming vector from the codebook, guaranteeing the radar operational requirements by mitigating inter-cell interference; and, 2) allocate time for communication, search, and tracking subframes to improve resource efficiency. The proposed strategy comprises a three-step algorithm that obtains the radar scan patterns and subframe allocation in order of task priority~\cite{Aubry2024}. The proposed solution works regardless of the priority order chosen, which depends on the operational requirements. Here, we set tracking to be the highest priority because of its time sensitivity, as delayed probing may cause large estimation errors; then, communication follows to ensure a minimal level of radar-communication coexistence under resource constraints; search has the lowest priority.

The operational requirements of each task are: for tracking, each \gls{bs} must probe all $N_\text{t}$ targets $\lfloor T_\text{f}R_\text{t} \rfloor$ times per frame, receiving echo signals with a target \gls{sinr} of $\bar{\Gamma}$; for communication, the network must achieve a throughput no less than $\bar{S}$; for search, each \gls{bs} must scan the entire cell area with detection probability at least $\bar{P}^\text{d}$. Thus, ($N_\text{t},R_\text{t},\bar{\Gamma},\bar{S},\bar{P}^\text{d})$ form the inputs to the task scheduling algorithm.

\subsection{Tracking Task Scheduling}\label{sec:tracking-scheduling}

To schedule the tracking tasks, the \glspl{bs}' scan pattern and the subframe duration required to probe all the tracked targets must be determined. To this end, we formulate the optimization of the tracking scan pattern to minimize the number of dwell intervals, subject to the radar \gls{sinr} constraint.

Let $\bm{\phi}^{(d)}\in\mathbb{C}^{N_\text{c}N_\text{a}}$ denote the beamformers set by all \glspl{bs} at the dwell interval $d$, and let $\Phi_\text{t}=\{\bm{\phi}^{(d)}\}_{d=1}^{D_\text{t}}$ denote the tracking scan pattern. For \gls{bs}~$i$, let $\Psi_i'\subseteq\Psi_i$ be the set of beamformers used to probe its $N_\text{t}$ tracked targets. We also define $\bm{\psi}_0$ as the null beamforming vector, indicating that a \gls{bs} does not transmit radar signals during a dwell interval.

Optimizing the beamformers w.r.t. the radar \gls{sinr} in \eqref{eq:sinr} requires knowledge of the interference terms, which depend on the locations of both detected and undetected targets. Since this is impractical, we introduce \emph{virtual scatterers} to estimate the \gls{sinr} of received radar signals.

\begin{definition}[Virtual Scatterers]\label{def:virtual-scatterers}
    For each cell and beamformer, a virtual scatterer is placed at the cell edge and aligned with a beamformer's half-power beamwidth. The set of virtual scatterers in cell~$i$ is denoted by $\mathcal{K}_i$, with $\bigcup_{i=1}^{N_\text{c}}\mathcal{K}_i=\mathcal{K}$ and $\mathcal{K}_i\cap\mathcal{K}_{j}=\emptyset$, $\forall i\neq j$. As such, $f_i:\Psi_i\rightarrow\mathcal{K}_i$ is a mapping of each beamformer to its corresponding virtual scatterer.
\end{definition}

Based on Def.~\ref{def:virtual-scatterers}, the \gls{sinr} of the radar signal received by \gls{bs}~$i$ is $\Gamma_{if_i(\bm{\phi}_i)}(\bm{\phi})$, and the tracking scan pattern can be optimized by solving the following problem:
\begin{subequations}
\label{eq:tracking-optimization}
\begin{align}
    \underset{\Phi_\text{t}}{\text{minimize}}\quad& D_\text{t} = |\Phi_\text{t}|\\
    \label{eq:tracking-sinr-constraint}
    \text{subject to}\quad& \Gamma_{if_i(\bm{\phi}_i^{(d)})}(\bm{\phi}^{(d)})\geq \bar{\Gamma},~ \forall i,d,\\
    \label{eq:tracking-completeness-constraint}
    \quad&\bigcup_{d=1}^{D_\text{t}}\{\bm{\phi}_i^{(d)}\}\backslash\{\bm{\psi}_0\}=\Psi_i',~ \forall i,\\
    \label{eq:tracking-uniqueness-constraint}
    \quad&\{\bm{\phi}_i^{(d)}\}\cap\{\bm{\phi}_i^{(d')}\}\backslash\{\bm{\psi}_0\}=\emptyset,~ \forall i,d\neq d'.
\end{align}
\end{subequations}
Constraint~\eqref{eq:tracking-sinr-constraint} enforces the radar target \gls{sinr}, while constraints~\eqref{eq:tracking-completeness-constraint} and~\eqref{eq:tracking-uniqueness-constraint} ensure that the scan pattern contains all beamformers required by the \glspl{bs}, and that each appears only once, except for $\bm{\psi}_0$.

For small network configurations, \eqref{eq:tracking-optimization} can be solved by enumeration. However, the number of potential solutions grows as $N_\text{t}^{N_\text{c}}!/(N_\text{t}^{N_\text{c}}-D_\text{t})!$ for a fixed $D_\text{t}$. To prove the effectiveness of the optimization in multi-cell scenarios, we propose a solution to \eqref{eq:tracking-optimization} for the scenario with $N_\text{c}=2$.\footnote{For $N_\text{c}>2$, the scan pattern optimization can be formulated as a graph coloring problem~\cite{Floudas2009}. In this formulation, the graph vertices represent the beamformers used by each \gls{bs}, while the edges connect pairs of beamformers whose simultaneous use violates the \gls{sinr} constraint in~\eqref{eq:tracking-sinr-constraint}.} This solution reformulates the original problem as an instance of the assignment problem, allowing efficient computation of the optimal scan pattern with standard solvers, even for large $N_\text{t}$.

\paragraph*{Solution of~\eqref{eq:tracking-optimization} for two cells}
Let the \emph{multisets}\footnote{Multisets consist of mathematical objects in which elements may appear with multiplicity greater than one.} with the indices of the beamformers of the two \glspl{bs} be initialized as $\mathcal{U} \gets \{u : \bm{\psi}_u \in \Psi_1'\}$ and $\mathcal{V} \gets \{v : \bm{\psi}_v \in \Psi_2'\}$. Then, we define a cost function that equals zero only when, using $\bm{\phi} = [\bm{\psi}_u^\transp~\bm{\psi}_v^\transp]^\transp$, the \gls{sinr} constraints at both \glspl{bs} are satisfied:
\begin{equation}
    \label{eq:cost}
    c(u,v) = \begin{cases}
        0, \text{ if } \Gamma_{1f_1(\bm{\psi}_u)}(\bm{\phi}) \text{ and } \Gamma_{2f_2(\bm{\psi}_v)}(\bm{\phi})\geq\bar{\Gamma},\\
        1, \text{ otherwise}.
    \end{cases}
\end{equation}
The problem of finding the matching beamformers indexed by $\mathcal{U}$ and $\mathcal{V}$ that minimize the total cost is formulated as
\begin{subequations}
\label{eq:assignment-problem}
\begin{align}
    \underset{\{x_{uv} \in \{0,1\}\}_{\forall u,v}}{\text{minimize}} \quad & C = \sum_{u\in\mathcal{U}} \sum_{v\in\mathcal{V}} c(u,v) x_{uv}\\
    \text{subject to}  \quad & \sum_{v\in\mathcal{V}} x_{uv} = 1,~ \forall u, \quad \sum_{u\in\mathcal{U}} x_{uv} = 1,~ \forall v.
\end{align}
\end{subequations}
This is the classical assignment problem, solvable in polynomial time by the Hungarian algorithm~\cite{kuhn1955hungarian}. If the optimal total cost $C^*=0$, then a matching of beamformers exists in which all the \gls{sinr} constraints are met. Otherwise, one of the \glspl{bs} must remain silent during one or more dwell intervals: thus, we update the multisets as $\mathcal{U}\gets\mathcal{U}+\{0\}$ and $\mathcal{V}\gets\mathcal{V}+\{0\}$, and solve \eqref{eq:assignment-problem} again. This process repeats until $C^*=0$. The optimized tracking scan pattern is then obtained from the solution $\{x_{uv}^*\}_{\forall u,v}$ of \eqref{eq:assignment-problem} as $\Phi_\text{t}^* = \{[\bm{\psi}_u^\transp~\bm{\psi}_v^\transp]^\transp : x_{uv}^* = 1,\forall u,v\}$. Accordingly, the number of required dwell intervals is $D_\text{t}^*=|\Phi_\text{t}^*|$, and from \eqref{eq:tracking-subframe}, the tracking subframe duration is $T_\text{t} = \lfloor T_\text{f}R_\text{t} \rfloor D_\text{t}^*T_\text{d}$. The overall procedure is summarized in Algorithm~\ref{alg:tracking-optimization}, which has worst-case complexity $O(|\mathcal{U}|^4)$ with efficient implementations of the Hungarian algorithm~\cite{Munkres1957}. Since the procedure starts with $\mathcal{U}$ and $\mathcal{V}$ containing only $\Psi_1'$ and $\Psi_2'$, and introduces silent dwell intervals gradually, $\Phi_\text{t}^*$ is always a feasible solution to~\eqref{eq:tracking-optimization}.


\begin{algorithm}[t]
    \caption{Scan pattern optimization for 2 cells.}
    \label{alg:tracking-optimization}
        \begin{algorithmic}[1]
            \renewcommand{\algorithmicrequire}{\textbf{input:}}
            \renewcommand{\algorithmicensure}{\textbf{output:}}  
            \REQUIRE Cost function $c$ and beamformers' indices $\mathcal{U},\mathcal{V}$
            \ENSURE Optimized scan pattern $\Phi^*$
            \STATE $(\{x_{uv}^*\}_{\forall u,v},C^*) \gets \mathrm{Hungarian\_algorithm}\{\text{\eqref{eq:assignment-problem}}\}$
            \IF{$C^* > 0$}
                \STATE $\mathcal{U}\gets\mathcal{U}+\{0\},\mathcal{V}\gets\mathcal{V}+\{0\}$ and go to line 1
            \ENDIF
            \RETURN $\Phi^* \gets \{[\bm{\psi}_u^\transp~\bm{\psi}_v^\transp]^\transp : x_{uv}^* = 1,\forall u,v\}$
        \end{algorithmic}
\end{algorithm}

\subsection{Communication Task Scheduling}\label{sec:communication-scheduling}

The communication tasks are scheduled to ensure the network achieves the required throughput $\bar{S}$. According to~\eqref{eq:throughput}, the communication subframe duration results in
\begin{equation}
    T_\text{c} = \frac{\bar{S}T_\text{f}}{W} \left( \sum_{i=1}^{N_\text{c}} \sum_{l\in\mathcal{L}_i} \log_2\left(1+\gamma_{il}\right) \right)^{-1}.
\end{equation}
If the available time resources are insufficient, \emph{i.e.}, $T_\text{f}-T_\text{t}<T_\text{c}$, the communication tasks are not scheduled. This occurs, \emph{e.g.}, when the tracking tasks occupy most of the frame to follow many fast-moving targets, or for high throughput requirements.

\subsection{Search Task Scheduling}\label{sec:search-scheduling}

For the search task, the \glspl{bs}' scan pattern is optimized to minimize the number of dwell intervals while satisfying a detection probability constraint. Letting $\Phi_\text{s}=\{\bm{\phi}^{(d)}\}_{d=1}^{D_\text{s}}$ denote the search scan pattern, the problem is formulated as
\begin{subequations}
\label{eq:search-optimization}
\begin{align}
    \underset{\Phi_\text{s}}{\text{minimize}}\quad& D_\text{s} = |\Phi_\text{s}|\\
    \label{eq:search-detection-constraint}
    \text{subject to}\quad& P_{if_i(\bm{\phi}_i^{(d)})}^\text{d}(\bm{\phi}^{(d)})\geq \bar{P}^\text{d},~ \forall i,d,\\
    \label{eq:search-completeness-constraint}
    \quad&\bigcup_{d=1}^{D_\text{s}}\{\bm{\phi}_i^{(d)}\}\backslash\{\bm{\psi}_0\}=\Psi_i,~ \forall i,\\
    \label{eq:search-uniqueness-constraint}
    \quad&\{\bm{\phi}_i^{(d)}\}\cap\{\bm{\phi}_i^{(d')}\}\backslash\{\bm{\psi}_0\}=\emptyset,~ \forall i,d\neq d'.
\end{align}
\end{subequations}
This formulation mirrors~\eqref{eq:tracking-optimization}, except that constraint~\eqref{eq:search-detection-constraint} replaces the \gls{sinr} constraint~\eqref{eq:tracking-sinr-constraint}. The problem reduces to an assignment problem that can be efficiently solved for $N_\text{c}=2$.

To solve \eqref{eq:search-optimization} for two cells, the cost function \eqref{eq:cost} is redefined to be zero when both $P_{1f_1(\bm{\psi}_u)}^\text{d}(\bm{\phi})$ and $P_{2f_2(\bm{\psi}_v)}^\text{d}(\bm{\phi}) \geq \bar{P}^\text{d}$. Then, running Algorithm~\ref{alg:tracking-optimization} with initialization $\mathcal{U} \gets \{u : \bm{\psi}_u \in \Psi_1\}$ and $\mathcal{V} \gets \{v : \bm{\psi}_v \in \Psi_2\}$ yields the optimized scan pattern $\Phi_\text{s}^* = \{[\bm{\psi}_u^\transp~\bm{\psi}_v^\transp]^\transp : x_{uv}^* = 1,\forall u,v\}$ with $D_\text{s}^* = |\Phi_\text{s}^*|$ dwell slots. The duration of the allocated search subframe is $T_\text{s} = \max\{T_\text{f}-T_\text{t}-T_\text{c},0\}$, and from \eqref{eq:search-rate}, the resulting search rate is $R_\text{s} = T_\text{s}/(D_\text{s}^*T_\text{d})$.

The three procedures described in Secs.~\ref{sec:tracking-scheduling}, \ref{sec:communication-scheduling}, and~\ref{sec:search-scheduling} form the proposed scheduling algorithm, used by the task scheduler to determine the subframe allocation and radar scan patterns.

\begin{table}[t]
    \centering
    \caption{Simulation Parameters}
    \begin{tabular}{cc}
        \toprule
        \textbf{Parameter} & \textbf{Value}\\
        \midrule
        Number of cells, radius & $N_\text{c}=2$, $R=100$~m\\
        Bandwidth, carrier wavelength & $W = 10$~MHz, $\lambda = 50$~mm\\
        Frame duration & $T_\text{f} = 1$~s\\
        \glspl{ue} per cell, transmit power & $|\mathcal{L}_i|=10$, $p_\text{c} = 23$~dBm\\
        Dwell time, number of pulses & $T_\text{d} = 13.3$~ms, $N_\text{p} = 20$\\
        Radar operational requirements & $\bar{P}^\text{d}=0.9$, $\bar{P}^\text{fa}=10^{-6}$, $\bar{\Gamma}=10$~dB\\
        \bottomrule
    \end{tabular}
    \label{tab:parameters}
\end{table}

%
%
\section{Numerical Results}

In this section, we provide numerical results to investigate the performance trade-offs of the proposed task scheduling strategy. The parameter values for the simulations are in Table~\ref{tab:parameters}. The transmit power budget for the radar tasks, $p_\text{r}$, is set to achieve the detection constraint ($\bar{P}^\text{d}$) or \gls{sinr} constraint~($\bar{\Gamma}$) for all virtual scatterers in the absence of interference. For the communication task, the \glspl{ue}' large-scale fading coefficients are calculated by $\log_{10}\beta_{jk}^i=-47.9-21\log_{10}r_{jk}^i$, where $r_{jk}^i$ is the distance from \gls{ue}~$k$ of cell~$j$ to \gls{bs}~$i$. The beamforming codebook is built using the phased array approach and \emph{Hamming weighting} with $N_\text{a}=29$, having an array power pattern $G(\theta;\bm{\phi}_i)$ given in~\cite[eq.~(3.22)]{VanTrees2002}, normalized to have a total transmit power of $p_\text{r}$; the half-power criterion is adopted to characterize the radar beamwidth. In this simplified setting, each beamforming vector corresponds to a single look direction.\footnote{The proposed algorithm for scan pattern optimization can be applied without loss of generalization to any beamforming codebook.}

Two baselines are adopted for comparison with the proposed scan patterns for search and tracking. In the \emph{in-phase scan pattern}, the \glspl{bs} beamformers point to the same relative directions at the same time, while, in the \emph{random scan pattern}, the \glspl{bs} select the beamformers randomly and independently. Only for the tracking task, the proposed scan pattern is compared with the \emph{orthogonal scan pattern}, where the \glspl{bs}' probes are orthogonal in the time domain, resulting in $D_\text{t}=N_\text{c}N_\text{t}$.

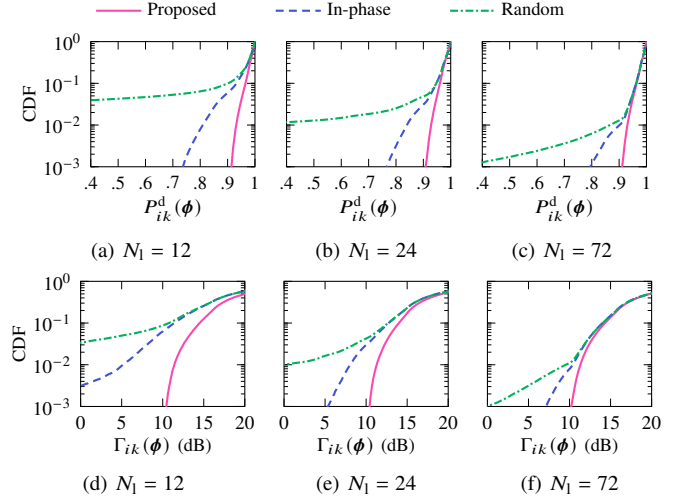
\begin{figure}[t]
    \centering
    \subfloat{%
    \begin{tikzpicture}
    \begin{axis}[
        width=0cm,
        height=0cm,
        axis line style={draw=none},
        tick style={draw=none},
        at={(0,0)},
        scale only axis,
        xmin=0,
        xmax=1,
        xtick={},
        ymin=0,
        ymax=1,
        ytick={},
        axis background/.style={fill=white},
        legend style={at={(0, 0)}, anchor=center, /tikz/every even column/.append style={column sep=2em}, legend columns=-1},
    ]
    \addlegendimage{draw=xgfs_normal6_pink, thick}
    \addlegendentry{Proposed}
    \addlegendimage{draw=xgfs_normal6_blue, thick, densely dashed}
    \addlegendentry{In-phase}
    \addlegendimage{draw=xgfs_normal6_green, thick, densely dashdotted}
    \addlegendentry{Random}
    \end{axis}
    \end{tikzpicture}
    }
    \\
    \setcounter{subfigure}{0}
    %
    %
    \subfloat[$N_\text{l}=12$]{%
    \begin{tikzpicture}
    \begin{axis}[
        ymode=log,
        width=0.425\columnwidth,
        ylabel={CDF},
        xlabel={$P_{ik}^\text{d}(\bm{\phi})$},
        xmin={0.4}, xmax={1}, 
        xtick={0.4,0.5,0.6,0.7,0.8,0.9,1}, xticklabels={.4,.5,.6,.7,.8,.9,1},
        ymin={1e-3}, ymax={1},  
    ]
        \addplot[thick,xgfs_normal6_pink] file{search_scan_pattern_Nl_12_opt.dat};
        \addplot[thick,xgfs_normal6_blue,densely dashed] file{search_scan_pattern_Nl_12_seq.dat};
        \addplot[thick,xgfs_normal6_green,densely dashdotted] file{search_scan_pattern_Nl_12_random.dat};
    \end{axis}
    \end{tikzpicture}}%
    %
    %
    \subfloat[$N_\text{l}=24$]{%
    \begin{tikzpicture}
    \begin{axis}[
        ymode=log,
        width=0.425\columnwidth,
        xlabel={$P_{ik}^\text{d}(\bm{\phi})$},
        xmin={0.4}, xmax={1}, 
        xtick={0.4,0.5,0.6,0.7,0.8,0.9,1}, xticklabels={.4,.5,.6,.7,.8,.9,1},
        ymin={1e-3}, ymax={1}, yticklabels=\empty,
    ]
        \addplot[thick,xgfs_normal6_pink] file{search_scan_pattern_Nl_24_opt.dat};
        \addplot[thick,xgfs_normal6_blue,densely dashed] file{search_scan_pattern_Nl_24_seq.dat};
        \addplot[thick,xgfs_normal6_green,densely dashdotted] file{search_scan_pattern_Nl_24_random.dat};
    \end{axis}
    \end{tikzpicture}}%
    %
    %
    \subfloat[$N_\text{l}=72$]{%
    \begin{tikzpicture}
    \begin{axis}[
        ymode=log,
        width=0.425\columnwidth,
        xlabel={$P_{ik}^\text{d}(\bm{\phi})$},
        xmin={0.4}, xmax={1}, 
        xtick={0.4,0.5,0.6,0.7,0.8,0.9,1}, xticklabels={.4,.5,.6,.7,.8,.9,1},
        ymin={1e-3}, ymax={1},
        yticklabels=\empty,
    ]
        \addplot[thick,xgfs_normal6_pink] file{search_scan_pattern_Nl_72_opt.dat};
        \addplot[thick,xgfs_normal6_blue,densely dashed] file{search_scan_pattern_Nl_72_seq.dat};
        \addplot[thick,xgfs_normal6_green,densely dashdotted] file{search_scan_pattern_Nl_72_random.dat};
    \end{axis}
    \end{tikzpicture}}\\
    %
    %
    \subfloat[$N_\text{l}=12$]{%
    \begin{tikzpicture}
    \begin{axis}[
        ymode=log,
        width=0.425\columnwidth,
        xlabel={$\Gamma_{ik}(\bm{\phi})$ (dB)},
        ylabel={CDF},
        xmin={0}, xmax={20}, xtick distance={5},
        ymin={1e-3}, ymax={1}
    ]
        \addplot[thick,xgfs_normal6_pink] file{tracking_scan_pattern_Nl_12_opt.dat};
        \addplot[thick,xgfs_normal6_blue,densely dashed] file{tracking_scan_pattern_Nl_12_seq.dat};
        \addplot[thick,xgfs_normal6_green,densely dashdotted] file{tracking_scan_pattern_Nl_12_random.dat};
    \end{axis}
    \end{tikzpicture}}
    %
    %
    \hfill
    \subfloat[$N_\text{l}=24$]{%
    \begin{tikzpicture}
    \begin{axis}[
        ymode=log,
        width=0.425\columnwidth,
        xlabel={$\Gamma_{ik}(\bm{\phi})$ (dB)},
        xmin={0}, xmax={20}, xtick distance={5},
        ymin={1e-3}, ymax={1}, yticklabels=\empty,
    ]
        \addplot[thick,xgfs_normal6_pink] file{tracking_scan_pattern_Nl_24_opt.dat};
        \addplot[thick,xgfs_normal6_blue,densely dashed] file{tracking_scan_pattern_Nl_24_seq.dat};
        \addplot[thick,xgfs_normal6_green,densely dashdotted] file{tracking_scan_pattern_Nl_24_random.dat};
    \end{axis}
    \end{tikzpicture}}
    %
    %
    \hfill
    \subfloat[$N_\text{l}=72$]{%
    \begin{tikzpicture}
    \begin{axis}[
        ymode=log,
        width=0.425\columnwidth,
        xlabel={$\Gamma_{ik}(\bm{\phi})$ (dB)},
        xmin={0}, xmax={20}, xtick distance={5},
        ymin={1e-3}, ymax={1}, yticklabels=\empty,
    ]
        \addplot[thick,xgfs_normal6_pink] file{tracking_scan_pattern_Nl_72_opt.dat};
        \addplot[thick,xgfs_normal6_blue,densely dashed] file{tracking_scan_pattern_Nl_72_seq.dat};
        \addplot[thick,xgfs_normal6_green,densely dashdotted] file{tracking_scan_pattern_Nl_72_random.dat};
    \end{axis}
    \end{tikzpicture}}
    \caption{CDF of the detection probability, $P_{ik}^\text{d}(\bm{\phi})$, given by the search scan pattern and radar \gls{sinr}, $\Gamma_{ik}(\bm{\phi)}$, given by the tracking scan pattern. The proposed search pattern when $N_\text{l}=12$ requires $D_\text{s}=13$, so each \gls{bs} needs to be turned off for one dwell interval. For all the other cases, $D_\text{s}=N_\text{l}$.}
    \label{fig:search}
\end{figure}

Fig.~\ref{fig:search} shows the search and tracking performance achieved by the different scan patterns and different $N_\text{l}$. The proposed pattern outperforms the baselines, guaranteeing $P^\text{d}_i(\bm{\phi})$ and $\Gamma_{ik}(\bm{\phi})$ to achieve the targets $\bar{P}^\text{d}$ and $\bar{\Gamma}$, respectively, with high reliability ($>0.999$). Conversely, the random pattern shows the worst performance, achieving values considerably lower than $\bar{P}^\text{d}$ and $\bar{\Gamma}$ with high probability, due to excessive inter-cell interference. The performance degradation of the baselines is reduced when increasing $N_\text{l}$, since having more look directions decreases the likelihood of the \glspl{bs} to simultaneously select beamforming vectors with high interference. However, even with $N_\text{l} = 72$, the baselines cannot guarantee the target performance with a reliability higher than $0.99$, which can only be achieved by the proposed patterns.

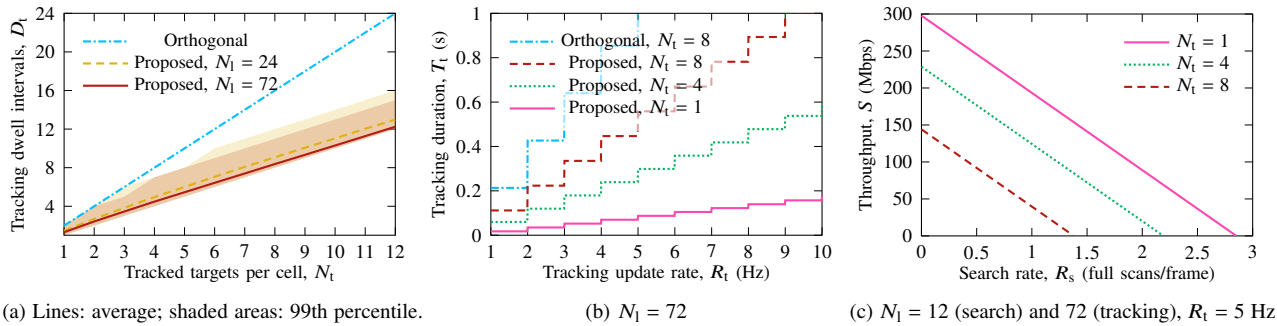
\begin{figure*}[b]
    \centering
    %
    %
    \subfloat[Lines: average; shaded areas: 99th percentile.]{%
    \begin{tikzpicture}
    \begin{axis}[
        width=0.33\linewidth,
        height=0.25\linewidth,
        xlabel={Tracked targets per cell, $N_\text{t}$},
        ylabel={Tracking dwell intervals, $D_\text{t}$},
        xmin={1}, xmax={12}, xtick distance={1},
        ymin={1}, ymax={24}, ytick distance={4},
        legend pos=north west,
        legend style={inner sep=3pt}
    ]
        \addplot[thick,xgfs_normal6_lightblue,densely dashdotted,domain={1:12}] {2*x};
        \addlegendentry{Orthogonal}
        
        \addplot[thick,xgfs_normal6_yellow,densely dashed] file{tracking_dwell_intervals_Nl_24.dat};
        \addlegendentry{Proposed, $N_\text{l}=24$}
        
        \addplot[thick,xgfs_normal6_red] file{tracking_dwell_intervals_Nl_72.dat};
        \addlegendentry{Proposed, $N_\text{l}=72$}

        \addplot[name path=Nl_72_upperb,draw=none] file{tracking_dwell_intervals_Nl_72_99thprctile.dat};
        \addplot[name path=Nl_72_lowerb,draw=none,domain={1:12}] {x};
        \addplot[xgfs_normal6_red,opacity=0.2] fill between[of=Nl_72_upperb and Nl_72_lowerb];

        \addplot[name path=Nl_24_upperb,draw=none] file{tracking_dwell_intervals_Nl_24_99thprctile.dat};
        \addplot[name path=Nl_24_lowerb,draw=none,domain={1:12}] {x};
        \addplot[xgfs_normal6_yellow,opacity=0.2] fill between[of=Nl_24_upperb and Nl_24_lowerb];
    \end{axis}
    \end{tikzpicture}
    \label{fig:tracking}
    }%
    %
    %
    \subfloat[$N_\text{l}=72$]{%
    \begin{tikzpicture}
    \begin{axis}[
        width=0.33\linewidth,
        height=0.25\linewidth,
        xlabel={Tracking update rate, $R_\text{t}$ (Hz)},
        ylabel={Tracking duration, $T_\text{t}$ (s)},
        xmin={1}, xmax={10}, xtick distance={1},
        ymin={0}, ymax={1}, ytick distance={0.2},
        legend pos=north west,
        legend style={inner sep=3pt,fill=white,fill opacity=0.6,text opacity=1}
    ]
        \addplot[thick,xgfs_normal6_lightblue,densely dashdotted] file{tracking_duration_Nt_8_orthogonal.dat};
        \addplot[thick,xgfs_normal6_red,densely dashed] file{tracking_duration_Nt_8_opt.dat};        
        \addplot[thick,xgfs_normal6_green,densely dotted] file{tracking_duration_Nt_4_opt.dat};
        \addplot[thick,xgfs_normal6_pink] file{tracking_duration_Nt_1_opt.dat};
        \legend{{Orthogonal, $N_\text{t}=8$}, {Proposed, $N_\text{t}=8$}, {Proposed, $N_\text{t}=4$}, {Proposed, $N_\text{t}=1$}}
    \end{axis}
    \end{tikzpicture}
    \label{fig:tracking-tradeoff}
    }%
    %
    %
    \subfloat[$N_\text{l}=12$ (search) and 72 (tracking), $R_\text{t}=5$~Hz]{%
    \begin{tikzpicture}
    \begin{axis}[
        width=0.33\linewidth,
        height=0.25\linewidth,
        xlabel={Search rate, $R_\text{s}$ (full scans/frame)},
        ylabel={Throughput, $S$ (Mbps)},
        xmin={0}, xmax={3}, xtick distance={0.5},
        ymin={0}, ymax={300}, ytick distance={50},
        legend pos=north east,
        legend style={inner sep=3pt}
    ]
        \addplot[thick,xgfs_normal6_pink] file{throughput_search_frequency_Nt_1.dat};
        \addplot[thick,xgfs_normal6_green,densely dotted] file{throughput_search_frequency_Nt_4.dat};
        \addplot[thick,xgfs_normal6_red,densely dashed] file{throughput_search_frequency_Nt_8.dat};
        \legend{$N_\text{t}=1$, $N_\text{t}=4$, $N_\text{t}=8$}
    \end{axis}
    \end{tikzpicture}
    \label{fig:communication-search-tradeoff}
    }%
    \caption{(a) Number of tracking dwell intervals required by the scan patterns vs. the number of tracked targets per cell. (b) Tracking subframe duration vs. the tracking update rate and number of tracked targets per cell. (c) Average communication throughput vs. the search rate and number of tracked targets per cell; the proposed patterns for search and tracking are adopted. All results are calculated over $10^4$ realizations.}
\end{figure*}

To investigate the tracking time performance, Fig.~\ref{fig:tracking} depicts the average number of dwell intervals (lines) and the 99th percentile (shaded areas) obtained for different numbers of tracked targets per cell and look directions. Compared to the orthogonal scan pattern, the proposed approach achieves a remarkable reduction in the number of required dwell intervals both for $N_\text{l}=24$ and $N_\text{l}=72$, especially for high $N_\text{t}$. Moreover, when $N_\text{l} = 72$, the scan pattern achieves the minimum possible length on average, having as many dwell intervals as targets. Thus, the proposed tracking pattern requires fewer time resources, improving efficiency by leaving more of the frame for search and communication tasks.

Fig.~\ref{fig:tracking-tradeoff} shows the duration of the tracking subframe achieved by different scan patterns against the tracking update rate and by varying $N_\text{t}$. As expected, $T_\text{t}$ grows with $R_\text{t}$ and $N_\text{t}$ due to the relation in eq.~\eqref{eq:tracking-subframe}, and the increase of $D_\text{t}$ with $N_\text{t}$---see Fig.~\ref{fig:tracking}. For $N_\text{t}=8$, it is noticeable that the duration of the proposed pattern presents a lower slope than the orthogonal one. In this case, the orthogonal pattern limits $R_\text{t}$ to 5~Hz before tracking occupies the entire frame of $1$s, while the proposed pattern supports values up to 9~Hz. Hence, the proposed approach allows each \gls{bs} to simultaneously track a larger number of moving targets whose positions change more often.

Fig.~\ref{fig:communication-search-tradeoff} shows the performance trade-offs of the tasks by showing the achievable region for communication throughput and search rate. The search and communication performance region degrades as $N_\text{t}$ and $R_\text{t}$ increase, as the longer tracking subframe, the fewer resources for the other tasks---see Fig.~\ref{fig:tracking-tradeoff}.

%
%
\section{Conclusion}

We proposed a scheduling strategy for multi-cell \gls{isac} networks optimizing communication and radar task allocation. Results showed that our approach outperforms conventional baselines by ensuring reliable radar detection and tracking with fewer resources. The framework provides an efficient means of coordinating \gls{isac} tasks at the \gls{mac} layer; future work will address scalability to larger networks and the impact of clutter.

%
%
\printbibliography

\balance

\end{document}